\documentclass[aps,prl,reprint,groupedaddress,nofootinbib]{revtex4-2}

\usepackage{amsmath}
\usepackage{graphicx}
\usepackage[colorlinks=true,citecolor=blue,urlcolor=blue]{hyperref}
\usepackage[capitalize]{cleveref}

\begin{document}

\title{EXAFS studies of the local environment of lead and selenium atoms \\
in PbTe$_{1-x}$Se$_x$ solid solution}

\author{A. I. Lebedev, I. A. Sluchinskaya, and V. N. Demin}
\affiliation{M. V. Lomonosov Moscow State University, 119899 Moscow, Russia}
\author{I. Munro}
\affiliation{Daresbury Laboratory, Warrington WA4 4AD, UK}


\begin{abstract}
EXAFS spectroscopy is used to study the local environment of lead and selenium atoms in
PbTe$_{1-x}$Se$_x$ solid solution. In addition to a bimodal distribution of bond lengths
in the first shell, unusually large Debye--Waller factors for the Pb--Pb interatomic
distances in the second shell and a substantial deviation of these distances from Vegard's
law are observed. Valence force field (VFF) calculations show that these observations are
due to the complex structure of the distribution function for Pb--Pb distances.
It is found that the number of Se--Se pairs in the second shell surpasses the statistical
value, which indicates that chemical factors play an important role in the structure of the
solid solution. The contribution of these chemical factors to the enthalpy of mixing of the solid
is estimated to be approximately 0.5 kcal/mole, which is comparable to the strain contribution.

Published in \texttt{Physics of the Solid State {\bf 41}, 1275 (1999).}
\end{abstract}

\maketitle


\section{1. INTRODUCTION}

Knowledge of the true structure of solid solutions, i.e., the local displacements of atoms from their
ideal positions in a lattice and the deviations of atomic distribution from the statistical one
is needed to understand the physical properties of these crystals. EXAFS spectroscopy is now widely
used for studying the structure of solid solutions.

EXAFS (extended x-ray-absorption fine-structure) studies of solid solutions of III--V and II--VI
semiconductors with a zinc blende structure~\cite{PhysRevB.28.7130,PhysRevB.31.7526,
SolidStateCommun.53.509,PhysRevB.41.8440,PhysRevB.42.11174,PhysRevB.48.8694} have revealed a
bimodal distribution of bond lengths in the first shell. This suggests that the fixed chemical bond
length approximation proposed by Bragg and Pauling is more suitable for describing the local
structure of solid solutions rather than the virtual crystal approximation. Up to now, solid
solutions of IV--VI semiconductors with the rocksalt NaCl structure have been studied by EXAFS only
to examine the locations of off-center impurities~\cite{PhysRevLett.59.2701,PhysicaB.158.578,
PhysRevB.46.11277,PhysRevB.55.14770}, no systematic investigations of the local structure have been
made for these solid solutions that contain no off-center impurities.

In this work, the results of an EXAFS study of the local structure of the PbTe$_{1-x}$Se$_x$ solid solution
are presented. This solid solution is of particular interest due to its use in creating IR optoelectronic
devices, specifically as a material for lattice-matched heterojunctions. The binary compounds PbTe and PbSe
form a continuous series of solid solutions across the entire composition range, with lattice parameters
of 6.460 and 6.126~{\AA} at 300~K, respectively. Their solid solution is characterized by a relatively
large difference in the bond lengths (approximately 5\%). Additionally, this solid solution is of
interest due to the fact that at $x \approx 0.25$ its lattice exhibits a maximum ``softening'' in respect
to the onset of the ferroelectricity when doped with Sn atoms~\cite{FizTverdTela.32.1780}.

\section{2. EXPERIMENTAL DETAILS}

Samples of PbTe$_{1-x}$Se$_x$ solid solution with $x = 0.1$, 0.25, 0.5, and 0.75 were prepared by
alloying stoichiometric PbTe and PbSe binary compounds in evacuated silica ampoules with subsequent
homogenizing annealing at 720$^\circ$C for 170~h. X-ray analysis found all the samples to be
single-phase. Just before the EXAFS measurements, the samples were ground, the obtained powders were
sifted through a mesh and deposited on a scotch tape. An optimum thickness of the absorbing layer
for taking spectra was obtained by multiple folding the tape (usually 8~layers).

The EXAFS measurements were carried out on the station 7.1 of the Daresbury synchrotron radiation
source (Great Britain) with an electron beam energy of 2~GeV and a maximum stored current of 230~mA.
Data were collected at the $K$ absorption edge of Se (12658~eV) and the $L_{\rm III}$ absorption
edge of Pb (13055~eV) at 80~K in the transmission mode. The synchrotron radiation was monochromatized
by a double-crystal Si(111) monochromator. The intensities of the incident radiation ($I_0$) and the
radiation transmitted through the sample ($I_t$) were measured by ionization chambers filled with
He+Ar mixtures, which absorb 20 and 80\% of the x-ray radiation, respectively. Contamination of the
incident radiation by higher harmonics can be neglected in the energy range used in the experiments.

The obtained transmission spectra $\mu x(E) = \ln(I_0/I_t)$ (where $E$ is the photon energy) were
analyzed in the usual way~\cite{RevModPhys.53.769}. After subtracting the pre-edge background, splines
were used to extract a smooth atomic part of absorption $\mu x_0(E)$, and then the EXAFS function,
$\chi = (\mu x - \mu x_ 0)/\mu x_0$, was calculated as a function of the photoelectron wave vector
$k = [2m(E - E_0)/\hbar^2]^{1/2}$. The energy origin, $E_0$, was determined by the position of the
inflection point on the absorption edge. The edge steps ranged from 0.08 to 1.4. At least two
independent measurements were taken for each sample.

When multiple scattering effects are neglected, information on the local environment of the central
atom, i.e., the distances $R_j$, coordination numbers $N_j$, and Debye--Waller factors $\sigma^2_j$
for each $j$th shell enters the EXAFS function as follows~\cite{RevModPhys.53.769}:
    \begin{equation}
    \begin{aligned}
    \chi(k) &= \frac{1}{k} \sum_j \frac{N_j S_0^2}{R_j^2} f( k) \exp \Big( - \frac{2R_j}{\lambda(k)} - 2k^2 \sigma^2_j \Big) \\
    &\times \sin[2kR_j + \psi(k)]. \\
    \end{aligned}
    \end{equation}
In addition to the structural parameters characterizing the local environment, this equation also
includes the amplitude $f(k)$, the sum of the backscattering and central atoms' phase shifts $\psi(k)$,
the photoelectron mean free path $\lambda(k)$, and the amplitude reduction factor $S_0^2$, which
accounts for many-electron and inelastic effects in the central and scattering atoms. The functions
$f(k)$, $\psi(k)$, and $\lambda(k)$ were calculated using the FEFF5 program~\cite{PhysRevB.44.4146}.

The information about two the nearest shells from the experimental $\chi(k)$ curves was isolated
using forward and inverse Fourier transforms with Hamming windows. The structural parameters $R_j$,
$N_j$, and $\sigma^2_j$ for the two shells as well as the energy origin correction $dE_0$ were determined
by minimizing the root-mean-square deviation between the experimental and calculated $k \chi(k)$
curves using a modified Levenberg--Marquardt algorithm. The number of fitting parameters (8--9)
was approximately half the number of independent data points ($2\Delta k \Delta R/\pi = {}$12--20,
where $\Delta k$ and $\Delta R$ are the range widths used for Fourier filtration in $k$ and $R$
spaces~\cite{RevModPhys.53.769}). The accuracy of the fitting parameters was estimated from the
covariance matrix, and the reported errors correspond to the standard deviation. To improve accuracy,
it was assumed that the $dE_0$ correction energy is the same for all shells and that the coordination
numbers correspond to the known coordination numbers in a fcc lattice.

\section{3. EXPERIMENTAL RESULTS}

\begin{figure}
\includegraphics{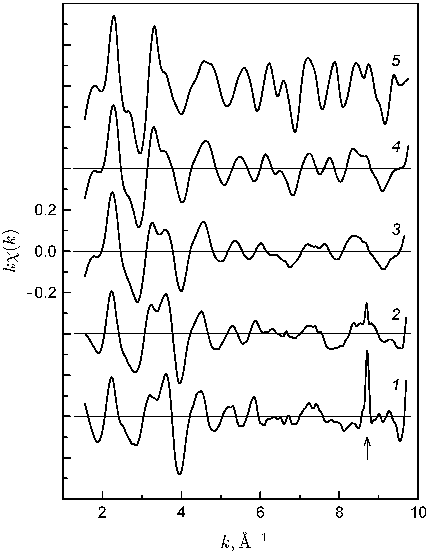}
\caption{\label{fig1}EXAFS $k \chi(k)$ spectra obtained at the $K$ absorption edge of Se for PbTe$_{1-x}$Se$_x$
samples. $x$: \emph{1}~--- 0.1, \emph{2}~--- 0.25, \emph{3}~--- 0.5, \emph{4}~--- 0.75, \emph{5}~--- 1. The
arrow denotes the location of a glitch.}
\end{figure}

\begin{figure}
\includegraphics{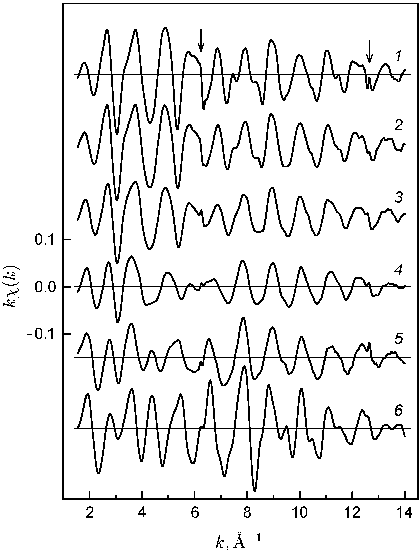}
\caption{\label{fig2}EXAFS $k \chi(k)$ spectra obtained at the $L_{\rm III}$ absorption edge of Pb for
PbTe$_{1-x}$Se$_x$ samples. $x$: \emph{1}~--- 0, \emph{2}~--- 0.1, \emph{3}~--- 0.25, \emph{4}~--- 0.5,
\emph{5}~--- 0.75, \emph{6}~--- 1. The arrows indicate the locations of glitches.}
\end{figure}

Fig.~\ref{fig1} displays typical $k \chi(k)$ curves obtained at the Se $K$ absorption edge for
PbTe$_{1-x}$Se$_x$ and PbSe. The maximum wave number for these data is limited by $k \approx 9.7$~{\AA}$^{-1}$
due to the proximity of the $L_{\rm III}$ absorption edge of Pb. Fig.~\ref{fig2} shows typical $k \chi(k)$
curves obtained at the Pb $L_{\rm III}$ absorption edge for PbTe$_{1-x}$Se$_x$, PbTe, and PbSe.
Sharp jumps on the curves marked by arrows at $k \approx 8.7$~{\AA}$^{-1}$ in Fig.~\ref{fig1} and
$k \approx 6.2$ and 12.7~{\AA}$^{-1}$ on Fig.~\ref{fig2}, the so-called glitches, are artifacts arising
from the monochromator and do not reflect the properties of the sample. After deglitching, the data were
processed as described in the previous section.

In analyzing data, we have limited ourselves to finding the parameters for two the nearest shells.
For the data obtained at the Pb absorption edge, we took into account that lead atoms are surrounded
by six atoms of two kinds (Se, Te) in the first shell. Each of the Pb--Te and Pb--Se atomic pairs is
characterized by its own set of structural parameters ($R_j$, $N_j$, $\sigma^2_j$). The second shell
consists of twelve Pb atoms that were thought to be located at the same distance.%
    \footnote{Studies of solid solutions with a zinc blende structure~\cite{PhysRevB.28.7130,
    PhysRevB.31.7526,SolidStateCommun.53.509} have shown that the
    atoms in the second shell can be located at different distances. The justification of the
    approximation we have made will be given below.}
In order to decrease the number of the fitting parameters, the relative contributions of the Pb--Te
and Pb--Se pairs into the theoretical $\chi(k)$ curves were taken according to the known chemical
composition of the samples, and the relative contributions of the first and second shells were taken
in accordance with their coordination numbers. The total number of the fitting parameters that describe
the local environment of lead was 8.

For the data obtained at the Se absorption edge, we took into account that the first shell of Se
always consists of six Pb atoms, while the second shell can contain either Se or Te atoms. It is
important to note that the numbers of Se and Te atoms in the second shell may differ from the statistical
values because of the manifestation of the short-range order in the solid solution. Thus, we analyzed
our data in both approximations, for random and nonrandom distributions of chalcogen atoms in the solid
solution.

In the case of a statistical distribution of chalcogen atoms, the problem was limited to finding
a set of eight parameters characterizing the local environment of Se atoms in the solid solution.%
    \footnote{To determine the parameters more accurately, additional constraints
    $R_{\rm Se-Te} = R_{\rm Se-Se}$ and $\sigma^2_{\rm Se-Te} = \sigma^2_{\rm Se-Se}$
    were introduced, and the Te and Se atoms were treated as a single shell. These constraints are
    justified by earlier studies of solid solutions which revealed a single-mode distribution of
    bond lengths in the sublattice, in which the isoelectronic substitution takes place~\cite{PhysRevB.28.7130,PhysRevB.48.8694},
    and by our VFF modeling of the structure of the PbTe$_{1-x}$Se$_x$ solid solution (see the
    next section).}

In order to take into account a possible nonrandom chalcogen distribution in the second shell,
an additional fitting parameter, $N_{\rm Se-Se}$, was introduced, which is the number of Se atoms
in the second shell of the central Se atom. We assumed that the total number of atoms in the second
shell is always 12.

\begin{table}
\footnotesize
\caption{\label{table1}Local concentration of Se--Se pairs in the PbTe$_{1-x}$Se$_x$ solid solution.}
\begin{ruledtabular}
\begin{tabular}{cccccc}
$x$ & 0.1 & 0.25 & 0.5 & 0.75 & 1.0 \\
\hline
$N_{\rm Se-Se}/12$ & 0.34$\pm$0.04 & 0.45$\pm$0.03 & 0.60$\pm$0.02 & 0.78$\pm$0.01 & 0.96$\pm$0.02 \\
\end{tabular}
\end{ruledtabular}
\end{table}

Taking the possibility of short-range order into account resulted in a considerably better
agreement between the experimental and theoretical $k \chi(k)$ curves compared to the model with
a random distribution of the chalcogen atoms. The obtained values of $N_{\rm Se-Se}$ (see Table~\ref{table1})
clearly indicate short-range order, namely, the preferential surrounding of
Se atoms in the second shell by atoms of the same type. This conclusion is consistent with that
of~\cite{ZMetallkd.70.541}, in which the thermodynamic properties of the PbTe--PbSe system were studied.

\begin{figure}
\includegraphics{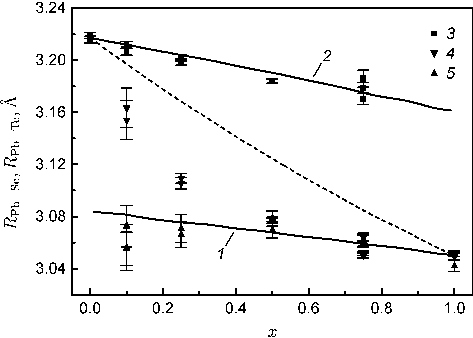}
\caption{\label{fig3}Pb--Se (\emph{1}) and Pb--Te (\emph{2}) bond lengths as a function of composition~$x$
of PbTe$_{1-x}$Se$_x$ solid solution. \emph{3}, \emph{4}, and \emph{5} are the results obtained from the
EXAFS data analysis for $R_{\rm Pb-Te}$, $R_{\rm Pb-Se}$, and $R_{\rm Se-Pb}$, respectively. Solid lines
and dashed line are mean Pb--Se, Pb--Se distances, and mean interatomic distance (calculated from the
lattice parameter) all obtained from the VFF modeling (Sec.~4).}
\end{figure}

\begin{figure}
\includegraphics{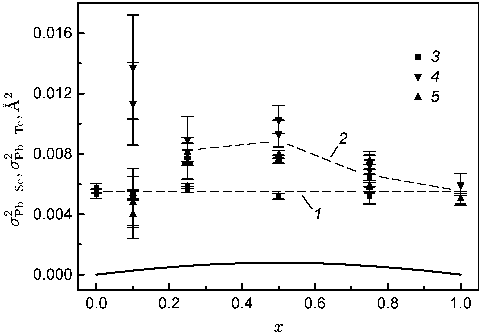}
\caption{\label{fig4}Debye--Waller factors for Pb--Te (\emph{1}) and Pb--Se (\emph{2}) bonds as functions
of the composition~$x$ of PbTe$_{1-x}$Se$_x$ solid solution. \emph{3}, \emph{4}, and \emph{5} are the
results of the EXAFS data analysis for Pb--Te, Se--Pb, and Pb--Se bonds, respectively. The solid line is
the static contribution to the Debye--Waller factor calculated using the VFF method (Sec. 4).}
\end{figure}

Interatomic distances and Debye--Waller factors for Pb--Te and Pb--Se bonds, derived from analysis
of both absorption edges, are plotted as a function of composition $x$ in Figs.~\ref{fig3} and \ref{fig4}.%
    \footnote{In the data analysis, we have taken into account systematic errors in the calculations
    of the theoretical scattering amplitudes and phases by the FEFF software. Distances obtained
    at the Pb edge were increased by 0.025~{\AA}~\cite{PhysRevB.55.14770}. A static correction of
    0.5~rad was added
    to the scattering phase for the Se edge data. These corrections yielded good agreement between
    EXAFS-derived distances and x-ray data for the reference PbTe and PbSe compounds.}
Despite the rather large errors in the determination of the Pb--Se distances in samples with low
selenium content, the distances obtained at both absorption edges agree well. As follows from
Fig.~\ref{fig3}, the interatomic bond length distribution in the PbTe$_{1-x}$Se$_x$ solid solution
demonstrates a clear bimodal character. The Debye--Waller factor's dependence on $x$ is different
for Pb--Te and Pb--Se bonds (Fig.~\ref{fig4}): while $\sigma^2_{\rm Pb-Te}$ is essentially
independent of the composition, $\sigma^2_{\rm Pb-Se}$ exhibits a maximum near $x = 0.5$.
Nevertheless, $\sigma^2$ remains below 0.01~{\AA}$^2$ for both bond types.

\begin{figure}
\includegraphics{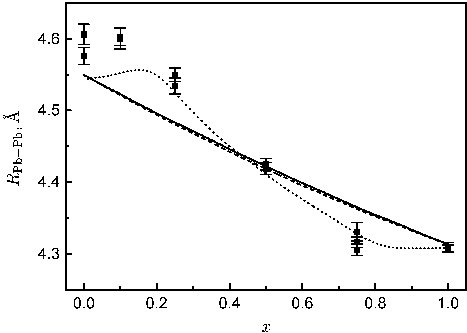}
\caption{\label{fig5}Pb--Pb interatomic distances as a function of $x$ in the PbTe$_{1-x}$Se$_x$
solid solution. The points are results of the EXAFS data analysis. The solid line is calculated
using the VFF method, the dashed line is the mean interatomic distance calculated from the lattice
parameter, and the dotted line presents the results of the EXAFS data analysis for simulated
$\chi(k)$ curves (equation~(\ref{eq4})).}
\end{figure}

\begin{figure}
\includegraphics{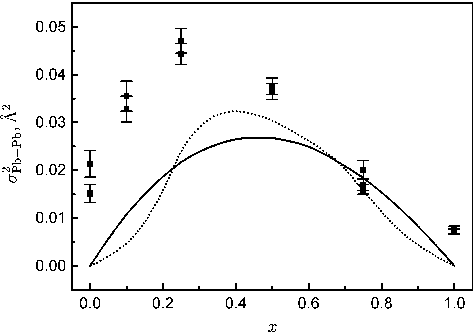}
\caption{\label{fig6}Debye--Waller factors for Pb--Pb pairs as a function of $x$ in the
PbTe$_{1-x}$Se$_x$ solid solution. The points are the results of the EXAFS data analysis. The
solid line is the result of the VFF calculations, and the dotted curve presents the results of
the EXAFS data analysis for simulated $\chi(k)$ curves (equation~(\ref{eq4})).}
\end{figure}

Figs.~\ref{fig5} and \ref{fig6} show the interatomic distances and Debye--Waller factors as a
function of $x$ for second-shell Pb--Pb atom pairs in the PbTe$_{1-x}$Se$_x$ solid solution.
Figs.~\ref{fig7} and \ref{fig8} present the same data for Se--chalcogen atom pairs, taking into
account the short-range order. Experimental results indicate a significant deviation from Vegard's
law for the experimental Pb--Pb distance dependence, whereas the Se--chalcogen pairs exhibit a better
agreement with Vegard's law.

The Debye--Waller factors for Pb--Pb and Se--chalcogen atom pairs exhibit similar qualitative
trends with respect to $x$, but their magnitudes differ. The Pb--Pb Debye--Waller factors show
a maximum near $x \approx 0.25$ and their values are unexpectedly large (up to 0.05~{\AA}$^2$),
substantially exceeding those of the Se--chalcogen pairs.

In order to explain the anomalously large Debye--Waller factors for Pb--Pb pairs in PbTe$_{1-x}$Se$_x$,
we supposed that the solid solution can transform into a microcoherent state, potentially
induced by mechanical treatment during the powder preparation, or the solid solution
decomposition due to insufficiently quick cooling of the samples after annealing. To check
this supposition, a second set of samples was prepared with the same compositions. Prior
to annealing, these samples were ground, and all the annealed powders were
quenched from 720$^\circ$C in cold water. The study of these samples at the Pb $L_{\rm III}$
absorption edge did not reveal any changes in the behavior of the Pb--Pb Debye--Waller factor
as a function of $x$ compared to samples of the first series. Moreover, a few samples of
the second series were additionally annealed at temperatures from 100 to 400$^\circ$C to
search for changes produced by a possible decomposition. No noticeable changes in the
Pb--Pb Debye--Waller factors were found.

To understand the reason for the unusual behavior of the Pb--Pb interatomic distances and
Debye--Waller factors (Figs.~\ref{fig5} and \ref{fig6}), we decided to simulate the static
distortions in the PbTe$_{1-x}$Se$_x$ solid solution using the valence force field (VFF)
method. This method has been used previously~\cite{SolidStateCommun.53.509,SolidStateCommun.55.413}
to model the structural distortions in semiconductor solid solutions with a zinc blende structure.

\section{4. VALENCE FORCE FIELD (VFF) MODELING}
\label{sec4}

The static distortions in the PbTe$_{1-x}$Se$_x$ solid solution were modeled within the
valence force field (VFF) approximation. For a given random distribution of Se and
Te atoms in one of the fcc sublattices, a set of atomic coordinates, \{${\bf r}_i$\},
for which the total strain energy associated with the bond stretching and deviation
of the angles between the bonds from 90$^\circ$,
    \begin{equation}
    \begin{aligned}
    U &= \frac{1}{2} \sum_i \Big[ \sum_{j = 1}^6 A_{s(i, j)} \Big( \frac{|{\bf r}_i - {\bf r}_j| - d_{s(i,j)}}{d_{s(i,j)}} \Big)^2 \\
      &+ B \sum_{(j,k) = 1}^{12} \Big( \frac{({\bf r}_i - {\bf r}_j) ({\bf r}_i - {\bf r}_k)}{|{\bf r}_i - {\bf r}_j| |{\bf r}_i - {\bf r}_k)|} \Big)^2 \Big], \\
    \end{aligned}
    \end{equation}
was minimized. In the first term, the sum runs over six nearest neighbors for each
lattice site. The indexes $s(i,j) = 1,2$ denote Pb--Se and Pb--Te pairs with atoms at sites
$i$ and $j$, $A_s$ is the stiffness constant for the corresponding bond, and $d_s$ is its
equilibrium length in the binary compound ($d_1 = 3.050$, $d_2 = 3.217$~{\AA} at 80~K). In
the second term, the sum runs over all 12 different angles between pairs of bonds for each
lattice site. Stiffness constants $A_1$ and $A_2$, associated with the elongation of Pb--Se
and Pb--Te bonds, respectively, were derived from published elastic moduli~\cite{LB-17f} for PbSe
and PbTe using the relation $A = (C_{11} + 2C_{12})/4$. The bending stiffness constants
were calculated as $B = C_{44}/32$, where $C_{44}$ represents the average shear modulus
of PbSe and PbTe.

Simulations were performed on $m \times m \times m$ lattices ($m = 10$--30) with
periodic boundary conditions, employing several realizations of random Se and Te distributions
within the chalcogen sublattice. Starting with ideal fcc lattice atomic positions, each
atom was displaced in a direction that minimized the strain energy, while maintaining
fixed positions of all other atoms. The step size for a single displacement was adaptively
reduced from 0.01 to 0.0002~{\AA}. After about 1000 steps per atom, the strain
energy $U$ converged, and the resulting atomic configuration was considered as the
equilibrium one. Distribution functions for all atomic pairs in the first and second shells
were then calculated from the obtained coordinate sets. As the periodic boundary conditions
required the lattice parameter to be fixed, the calculations were done for a set of lattice
parameters. Its value corresponding to the minimum $U$ was subsequently adopted as the lattice
parameter.

Numerical simulations performed on lattices of different size showed that the excess energy
due to the finite-size effects scales approximately as $m^{-3}$ and that for $m > 16$, these
effects can be neglected.

The calculated mean Pb--Se and Pb--Te bond lengths and their dispersions (static Debye--Waller
factors) are represented by solid lines in Figs.~\ref{fig3} and \ref{fig4}. As can be expected,
a clear bimodal distribution is observed for the distances in the first shell. Let us define
the relaxation parameter $\epsilon'$ for the $A$--$C$ bond as the variation in its length in
two limiting compositions of the $AB_{1-x}C_x$ solid solution, normalized by the total change
in the mean interatomic distance,
    \begin{equation}
    \epsilon' = \frac{R_{AC}^0 - R_{AC}[AB]}{R_{AB}^0 - R_{AC}^0},
    \end{equation}
where $R_{AB}^0$ and $R_{AC}^0$ are the bond lengths in binary $AB$ and $AC$ compounds, and
$R_{AC}[AB]$ is the $A$--$C$ bond length for the impurity $C$ in the $AB$ compound at infinite
dilution. For the Pb--Te and Pb--Se bonds, the corresponding values are 0.35 and 0.21, respectively.
The difference between the two values of $\epsilon'$ is due to an appreciable difference in the
bond-stretching constants $A_1$ and $A_2$ (about 1.5 times). The value of $\epsilon'$ calculated
for the Pb--Te bond agrees well with the experimental data ($\epsilon' \approx 0.36$). The
comparison for the Pb--Se bond is complicated by larger experimental errors. It is noteworthy
that the calculated dependence of the mean interatomic distance on $x$ (dashed line in Fig.~\ref{fig3})
displays a noticeable negative bowing compared to Vegard's law (by about 0.009~{\AA}).

\begin{figure}
\includegraphics{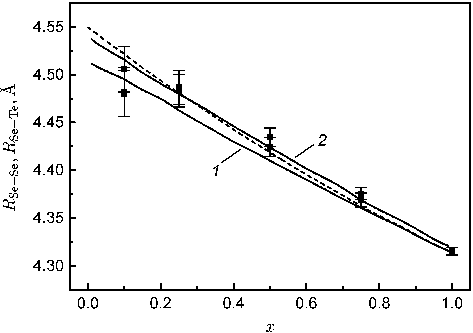}
\caption{\label{fig7}Se--Se (\emph{1}) and Se--Te (\emph{2}) interatomic distances as functions
of $x$ in the PbTe$_{1-x}$Se$_x$ solid solution. The points are the results of the EXAFS data
analysis. The solid lines are the results of the VFF calculations, and the dashed line is the mean
interatomic distance calculated from the lattice parameter.}
\end{figure}

\begin{figure}
\includegraphics{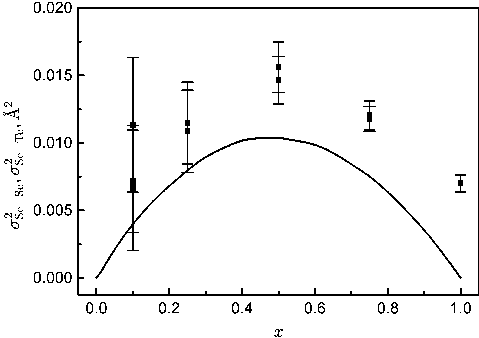}
\caption{\label{fig8}Debye--Waller factors for Se--Se and Se--Te pairs as functions of $x$ in
the PbTe$_{1-x}$Se$_x$ solid solution. The points are the results of the EXAFS data analysis.
The solid line is calculated using the VFF method.}
\end{figure}

Mean interatomic distances and their dispersions calculated for Pb--Pb, Se--Se, and Se--Te pairs
in the second shell are presented in \cref{fig5,fig6,fig7,fig8}. Very large static Debye--Waller
factors in the second shell compared to those in the first shell suggest that the lattice adapts
to the existence of different bond lengths by local rotations. The results of VFF simulations for
the Se--chalcogen pairs agree well with experimental data, whereas the agreement for the Pb--Pb
pairs is not good. The reasons for this disagreement will be discussed in detail in the next section.

Our calculations show that the mean Se--Se distance in the solid solution is only 0.01--0.02~{\AA}
shorter than the Se--Te distance (see Fig.~\ref{fig7}), and their Debye--Waller factors are
nearly equal ($\sigma^2_{\rm Se-Se} \approx \sigma^2_{\rm Se-Te}$). This justifies treating the
chalcogen atoms as a unified single shell in our data analysis at the Se edge.

In summary, valence force field modeling of distortions in the PbTe$_{1-x}$Se$_x$ solid solution
has predicted inevitably strong static distortions of Pb--Pb distances and demonstrated
a qualitative agreement between the calculations and experiment for the first shell and for the
Se--chalcogen pairs in the second shell. However, the modeling has not provided an explanation
for the discrepancy between the experimentally determined Pb--Pb distances and calculated data,
potentially due to the inadequacy of the single-mode approximation for Pb--Pb distances used in
the data analysis.

\section{5. DISCUSSION}

We start our discussion with a question: Is a single-mode approximation sufficient to describe
interatomic distances in the second Pb--Pb shell of the PbTe$_{1-x}$Se$_x$ solid solution?

It has been proposed~\cite{PhysRevB.28.7130,PhysRevB.31.6903} that the number of modes in the
distribution of the $A$--$A$ distances
in an $AB_{1-x}C_x$ solid solution is determined by the number of different ways that link two
$A$ atoms, i.e. the distribution should be bimodal for lattices with a zinc blende structure
and trimodal for NaCl structure.

That is why we initiated our data analysis by seeking parameters that approximate the distribution
function as a sum of a few Gaussians. Unfortunately, a large Debye--Waller factor and a limited $k$
range of EXAFS oscillations did not enable us to determine the parameters for a trimodal approximation
(the solution was not stable). For the bimodal approximation, we succeeded in getting reasonable
parameters for compositions close to binary compounds; for the sample with $x = 0.5$, the distance
to one of the components became unphysically large ($\sim$4.7~{\AA}). Since neither the trimodal
approximation nor the bimodal one could be used, for all the samples we had to analyze our data
using the single-mode approximation.

\begin{figure}
\includegraphics{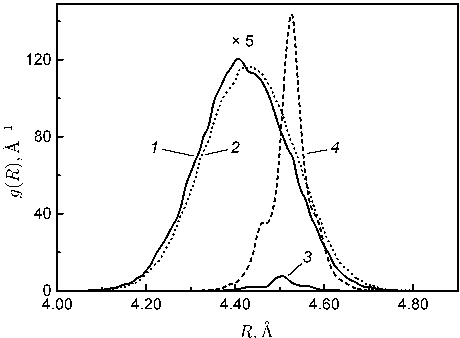}
\caption{\label{fig9}Se--Se (\emph{1}, \emph{3}) and Se--Te (\emph{2}, \emph{4}) pair distribution
functions calculated using the VFF method for the PbTe$_{1-x}$Se$_x$ solid solution with $x = 0.5$
(\emph{1}, \emph{2}) and $x = 0.05$ (\emph{3}, \emph{4}).}
\end{figure}

\begin{figure}
\includegraphics{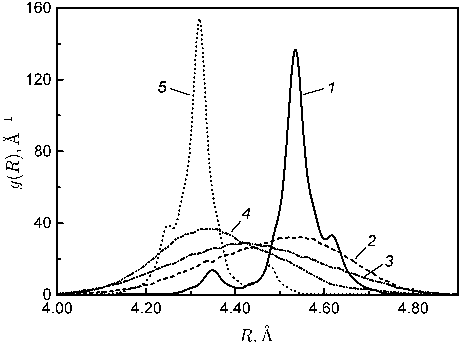}
\caption{\label{fig10}Pb--Pb pair distribution functions calculated using the VFF method for
PbTe$_{1-x}$Se$_x$ solid solution with $x$: \emph{1}~--- 0.05, \emph{2}~--- 0.25, \emph{3}~--- 0.5,
\emph{4}~--- 0.75, \emph{5}~--- 0.95.}
\end{figure}

Let us consider the distribution functions $g(R)$ for Se--Se, Se--Te, and Pb--Pb pairs, calculated
using the VFF method (Figs.~\ref{fig9} and \ref{fig10}). It is seen that for $x \approx 0$ and
$x \approx 1$, the distribution functions display a fine structure, which is smeared out as
$x \to 0.5$. However, the distribution functions for Se--chalcogen and Pb--Pb pairs are very
different. For the Se environment, the mean Se--Se and Se--Te interatomic distances are close,
and the distance between the corresponding components in the fine structure (Fig.~\ref{fig9})
is small, typically about the amplitude of thermal vibrations at 80~K. These results confirm the common
pattern~\cite{PhysRevB.28.7130,PhysRevB.48.8694} that the distribution function for a pair of atoms that
substitute each other in the solid solution $AB_{1-x}C_x$ can be treated as single-mode.

As follows from Fig.~\ref{fig10}, the distribution function for the $A$--$A$ pair in our solid solution
with an NaCl structure cannot be described by the trimodal approximation. Nevertheless,
sometimes the fine structure components can be divided tentatively into two groups whose centers
of gravity differ significantly. Therefore, it appears that when the smearing is not too large,
it is possible to approximate the distribution function by a sum of two Gaussians.

In order to check whether the Pb--Pb distribution functions obtained from the VFF calculations
are consistent with the experimental data, we calculated $\chi(k)$ curves using formula
    \begin{equation}
    \begin{aligned}
    \chi(k ) &= \frac{S_0^2}{k} f(k) \int \frac{{g(R)}}{R^2} \exp \Big( - \frac{2R}{\lambda(k)} \Big) \\
    &\times \sin [ 2kR + \psi(k) ] \, dR. \\
    \end{aligned}
    \label{eq4}
    \end{equation}
Then, the calculated curves were analyzed in a way similar to that used for the experimental
$\chi(k)$ curves (one distance for the Pb--Pb pair). The interatomic distances and static
Debye--Waller factors obtained in this way are plotted with dotted lines in Figs.~\ref{fig5} and
\ref{fig6}. It is seen that the obtained data are in good agreement with the experimental data,
i.e., the distribution functions calculated using the VFF method provide a qualitatively accurate
description of the real structure of the solid solution.

Our results, therefore, show that in the PbTe$_{1-x}$Se$_x$ solid solution, the distortions
in the second shell of chalcogen sublattice (where substitution occurs) can be described
with sufficient accuracy using a single-mode approximation, whereas for the Pb sublattice
(not subject to substitution), it is difficult to find a good working approximation, and the
best solution is to use distribution functions calculated using the VFF method.

The experimentally measured Debye--Waller factors are known to consist of static and dynamic
terms that correspond to static distortions of the structure and thermal lattice vibrations,
respectively. The VFF calculations give only the static part of the Debye--Waller factor.
Thus, the difference between the experimental and calculated Debye--Waller factors can be
attributed to thermal vibrations. For the first shell (Fig.~\ref{fig4}), the dynamic contribution
to the Debye--Waller factor is an order of magnitude greater than the static contribution, even
at 80~K. The difference in the factors for Pb--Te and Pb--Se bonds is probably related to
the appreciable difference in masses of Te and Se atoms. A more detailed analysis of these
data is difficult because of the bimodal character of the phonon spectrum of PbTe$_{1-x}$Se$_x$
solid solution~\cite{PhysStatusSolidiB.95.179}. The contribution of thermal vibrations to the
Debye--Waller factor
for Se--chalcogen pairs (Fig.~\ref{fig8})  in the second shell is a little bit higher than for the
first shell (atomic motion in the second shell is less correlated), and within the accuracy of
the experiment, it does not depend on $x$. However, as follows from Fig.~\ref{fig6}, for Pb--Pb
pairs the dynamic contribution depends on $x$; the maximum amplitude of thermal vibrations occurs
at $x \approx 0.25$. This result qualitatively agrees with the data of our paper~\cite{FizTverdTela.32.1780},
where it was shown that the structure of the PbTe$_{1-x}$Se$_x$ solid solution becomes very ``soft''
at $x \approx 0.25$, enabling us to observe the onset of the phase transition when a part of
Pb atoms are substituted by tin.

We now consider the relaxation of bonds in the solid solution. As shown above, the parameter
$\epsilon'$ for the PbTe$_{1-x}$Se$_x$ solid solution was $\approx$0.36, i.e., intermediate
between the values observed previously in III--V and II--VI semiconductors with a zinc blende
structure (0.05--0.24~\cite{PhysRevB.42.11174}) and those in alkali halide crystals with an
NaCl structure ($\approx$0.5~\cite{PhysRevB.31.6903}).

It is obvious that one of the main factors resulting in changes in the bond lengths is the
appearance of elastic strain caused by the difference in size of the substituted $B$ and
$C$ atoms. As follows from simple considerations~\cite{PhysRevB.31.1139}, the $\epsilon'$ value
is determined by the crystal structure (mutual position of the atoms in the lattice), so that
only the data for the same structures can be compared. Neglecting the bond bending and atomic
displacements in the second and more distant shells, the authors of~\cite{PhysRevB.31.1139}
obtained the value of $\epsilon' = 0.25$
for a zinc blende structure and $\epsilon' = 0.5$ for an NaCl structure, assuming equal
stiffness for $A$--$B$ and $A$--$C$ bonds. More realistic VFF calculations performed in this
work, which took the influence of all the factors listed above into account, yielded
$\epsilon' = 0.35$ and 0.21 for the two bonds in the PbTe$_{1-x}$Se$_x$ solid solution with
an NaCl structure, which is in fair agreement with experiment.

However, the strain contribution alone appears not to be sufficient to explain all the experimental
data. Our attempt to calculate $\epsilon'$ using the VFF method for the alkali halide RbBr$_{1-x}$I$_x$
and K$_{1-x}$Rb$_x$Br crystals with an NaCl structure studied in~\cite{PhysRevB.31.6903} yielded
$\epsilon'$ values close to those obtained for PbTe$_{1-x}$Se$_x$, but differing substantially
from experiment. A large difference between the experimental values of $\epsilon'$ for III--V and
II--VI semiconductors with a zinc blende structure and those predicted by the strain model has
also been noticed in~\cite{PhysRevB.42.11174}. In our opinion, these discrepancies are associated
with another group of factors which can influence the local structure, namely, the so-called
chemical factors, which take into account the individual properties of interacting atoms (their
valence, electronegativity), and type of chemical bonding.

These factors are also manifested in thermodynamic properties; it is known that the enthalpy
of mixing of a solid solution, $\Delta H_m$, consists of two components, the strain energy and
chemical contribution. The strain energy $U$ calculated using the VFF method is actually the strain
contribution to $\Delta H_m$. In III--V and II--VI semiconductors with a zinc blende structure,
the strain energy agrees well with the experimental mixing enthalpy~\cite{JPhysCondensMatter.6.4437}.
Unfortunately, the
available published data on the enthalpy of mixing of the PbTe$_{1-x}$Se$_x$ solid solution
($\Delta H_m \le 0.2$~kcal/mole at $x = 0.5$) are not very accurate. Although $\Delta H_m$ is
positive, thus indicating a tendency to decomposition of the solid solution, the difference
between the calculated strain contribution to $\Delta H_m$ ($U = 0.38$~kcal/mole) and the
measured $\Delta H_m$ value may be due to the presence of chemical factors.

Individual members of this group of factors may either increase or decrease the $\epsilon'$
values (the chemical bond length). We suppose that in the solid solution under study, the
effect of different factors are compensated, and the agreement between the calculated and
experimental $\epsilon'$ remains good.

From the local concentration of Se--Se pairs (short-range order) obtained in our experiment,
the contribution of chemical factors into the mixing enthalpy can be estimated. We define the
freezing temperature $T_f$ as a temperature at which the interchange of the two nearest atoms
takes one second (it corresponds to the diffusion coefficient of $D = 2 \times10^{-15}$~cm$^2$/s).
According to the published data on the temperature dependence of the diffusion coefficient~\cite{LB-17d}, 
the estimated value of $T_f$ is 540~K. Using the approximation of pairwise atomic interaction,
we calculated the difference between the energy of Se--Te pair and the average energy of
Se--Se and Se--Te pairs. This value appeared to be $\Delta E \approx 0.5$~kcal/mole (0.02~eV)
for the sample with $x = 0.5$. This value is actually an estimate of the \emph{chemical
contribution} into the mixing enthalpy: the atoms in the second shell do not interact
elastically, and so there is no elastic contribution to this value. The comparison of the
$\Delta E$ and $U$ values shows that the chemical and elastic contributions into the mixing
enthalpy in the PbTe$_{1-x}$Se$_x$ solid solution have the same order of magnitude.

However, the use of a pairwise atomic interaction approximation is not fully correct, since
the atoms in the second shell do not interact directly (either chemically or elastically).
So, the appearance of the short-range order is apparently due to some more complex interactions
involving three or more atoms. In our experiments, the largest deviations from the random
distribution of chalcogen atoms were observed at low Se concentrations and was manifested as
the increased number of Se--Pb--Se configurations with Pb--Se bonds oriented at 90$^\circ$.
This suggests that there is an appreciable difference in the energies of configurations in
which the like atoms belong to the same or to different $p$ orbitals. These configurations
can be schematically shown in the following way:

$$
\begin{array}{cccccccccccc}
{\rm Se} &          &          &          & {\rm Te} & \qquad & {\rm Se} &          &          &        & {\rm Te} \\
         & \nwarrow &          & \nearrow &          &        &          & \nwarrow &          & \nearrow   & \\
         &          & {\rm Pb} &          &          &        &          &          & {\rm Pb} &        & \\
         & \swarrow &          & \searrow &          &        &          & \swarrow &          & \searrow   & \\
{\rm Te} &          &          &          & {\rm Se} &        & {\rm Se} &          &          &        & {\rm Te} \\
\end{array}
$$

It cannot be excluded that the cause of such a behavior is associated with unsaturated character of
chemical bonds in IV--VI semiconductors.

In conclusion, the use of the VFF method to calculate the bond lengths and pair correlation
functions for the first and second shells enables us to describe qualitatively the distortions
of the structure in the PbTe$_{1-x}$Se$_x$ solid solution. However, this model does not predict
the scatter in the experimental $\epsilon'$ values, for which the model predicts close values
of this parameter, as well as the appearance of the short-range order. This evidences that along
with elastic interactions, the influence of chemical factors should be taken into account to
describe the properties of solid solutions.

\begin{acknowledgments}
This work was partially supported by the Russian Foundation for Basic Research (Grant No. 95-02-04644).
\end{acknowledgments}


%

\end{document}